\documentclass[twocolumn,trackchanges]{aastex701}
\shorttitle{VELC Spectroscopy of a Fast CME}
\shortauthors{Tripathi, Samanta, \& Joshi}
\graphicspath{{./}{figures/}}

\usepackage{graphicx}        
\usepackage{amssymb}         
\usepackage{color}           
\usepackage{breakurl}        
\usepackage{caption}

\usepackage[percent]{overpic}
\usepackage{amsmath}
\usepackage{caption} 
\usepackage{array}           
\usepackage{booktabs}        
\usepackage{multirow}        
\usepackage{bigstrut}        
\usepackage{mathrsfs}
\usepackage[utf8]{inputenc} 

\usepackage{xcolor}

\chardef\us=`\_
\begin{document}

\title{Spectroscopic Signatures of Nonuniform Lateral Expansion of a Coronal Mass Ejection in the Low Corona with ADITYA-L1/VELC}
\author[0009-0006-7404-4235]{Saurabh Tripathi}
\affiliation{Indian Institute of Astrophysics, Koramangala, Bangalore, 560034, India}
\affiliation{Pondicherry University, 605014, Puducherry, India}
\email{saurabh.tripathi@iiap.res.in}  

\author[orcid=0000-0002-9667-6392]{Tanmoy Samanta} 
\affiliation{Indian Institute of Astrophysics, Koramangala, Bangalore, 560034, India}
\affiliation{Pondicherry University, 605014, Puducherry, India}
\email[show]{tanmoy.samanta@iiap.res.in}

\author[orcid=0000-0003-0585-7030]{Jayant Joshi}
\affiliation{Indian Institute of Astrophysics, Koramangala, Bangalore, 560034, India}
\affiliation{Pondicherry University, 605014, Puducherry, India}
\email{jayant.joshi@iiap.res.in}

\begin{abstract}
Coronal mass ejections (CMEs) undergo rapid evolution in the low corona ($<2R_\odot$), yet their three-dimensional kinematics remain poorly constrained. We present a spectroscopic and imaging analysis
of a CME observed on 2024 August 5 using the Visible Emission Line Coronagraph (VELC) onboard \textit{ADITYA-L1} and \textit{AIA/SDO}. The VELC Fe\,{\sc xiv} 5303~\AA\ spectra captured the erupting CME as it crossed the western slit. Space–time maps from single-Gaussian fits reveal coronal dimming and enhanced Doppler velocities during the eruption. Detailed spectral analysis reveals strongly Doppler-shifted secondary components associated with the CME front. These components appear as two spatially separated emission branches on opposite sides of the rest wavelength, producing a distinct bifurcated spectral morphology superposed on the nearly stationary, reduced-intensity coronal emission. Double-Gaussian fitting yields line-of-sight (LOS) plasma velocities of up to $218~\mathrm{km\,~s^{-1}}$, likely a lower limit as the secondary component reaches the edge of spectral window. AIA 211~\AA\ observations measure a projected radial speed of $1135$~km~s$^{-1}$, larger than the plane-of-sky (POS) lateral speed of $692$~km~s$^{-1}$ along the slit. To interpret the observed CME evolution and spectral structure, we developed a simplified three-dimensional geometric-kinematic model of an expanding, axially tilted CME shell.  Synthetic spectra from the model show that an inclined CME undergoing asymmetric lateral expansion can produce a bifurcated spectral morphology similar to that observed, with velocity components projected in the radial, POS-lateral, and LOS-lateral directions. Our results demonstrate that combining spectroscopy and EUV imaging constrains different components of CME kinematics during early evolution.
\end{abstract}


\section{Introduction} 
\noindent 
Coronal mass ejections (CMEs) are large-scale eruptions of magnetized plasma, often associated with solar flares and the release of coronal magnetic energy \citep{1990GeoRL..17..901G, 2012LRSP....9....3W, 2011LRSP....8....6S,10.1098/rsta.2018.0094, 10.1063/1.4993929}. Their initial acceleration and structural evolution in the low corona influence their subsequent propagation in the heliosphere \citep{2004ASSL..317..201G,2011LRSP....8....1C,10.3389/fspas.2021.651527,10.3389/fspas.2023.1264226}. However, observations of this early phase remain limited because coronagraphs have limited coverage of the low corona.

White-light observations from LASCO coronagraph onboard the Solar and Heliospheric Observatory (SOHO) and COR2 coronagraph onboard STEREO mainly track CME structures above about $2~R_\odot$ \citep{1995SoPh..162..357B, 2008SSRv..136...67H, 2012LRSP....9....3W}. Multi-viewpoint observations have enabled reconstruction of the three-dimensional CME morphology and propagation direction using geometric models such as the Graduated Cylindrical Shell (GCS) model \citep{2006ApJ...652..763T, 2009SoPh..256..111T, 2011ApJS..194...33T}. These observations have also shown that CME expansion can differ between the radial and lateral directions \citep{2020ApJ...899....6M,2026ApJ..1008..137A}.

The early evolution of CMEs in the low corona is primarily observed using EUV imagers, such as Extreme ultraviolet Imaging Telescope (EIT) onboard SOHO and Atmospheric Imaging Assembly (AIA) \citep{2012SoPh..275...17L} onboard the Solar Dynamics Observatory (SDO), which capture the plane-of-sky (POS) evolution of erupting structures and associated phenomena such as EUV waves and coronal dimmings \citep{1998ApJ...498L.179G,1999ApJ...520L.139Z, 2010ApJ...724L.188P,2010SoPh..265....5M,Ma_2011, 2011ApJ...733L..25K, 2012SoPh..281..187P, 2012ApJ...745L...5C, 2012ApJ...753...52L, 2014SoPh..289.3233L, 2015LRSP...12....3W,2017JSWSC...7A..32K,2020SoPh..295..124F, 2025LRSP...22....2V}. Spectroscopic observations from Ultraviolet Coronagraph Spectrometer (UVCS) onboard SOHO \citep{1995SoPh..162..313K} and EUV Imaging Spectrometer (EIS) onboard Hinode \citep{2007SoPh..243...19C} have shown that CMEs contain multiple plasma components with different line-of-sight (LOS) velocities \citep{Raymond_2000,Landi_2010,2012ApJ...748..106T}. UVCS observations of fast, flare-associated CMEs have revealed large Doppler shifts associated with rapid expansion and high-temperature plasma during eruptions \citep{2003ApJ...597.1106R}. Such observations provide information on CME motions along the line of sight that cannot be obtained from imaging alone. Ground-based spectroscopic observations and forward-modeling studies have further demonstrated the potential of coronal emission-line spectroscopy to constrain the kinematics, thermodynamic properties, and magnetic structure of CMEs in the low corona \citep{2013SoPh..288..637T,2023SoPh..298..112L,2025ApJ...980...30W,2025ApJ...984..141C}.

The Visible Emission Line Coronagraph (VELC) onboard ADITYA-L1 \citep{2019AdSpR..64.1455S, 2024AdSpR..74..547P, 
2024ApJ...976L...6R,
2025SoPh..300...66S, 2025ApJ...994..182P,
2025ApJ...983..171M} provides spectroscopic observations of the low corona. Its multi-slit design enables high-cadence observations of the Fe\,{\sc xiv} 5303~\AA\ green line, which traces hot ($\sim1.8$~MK) coronal plasma at heights of 1.05--1.5~$R_\odot$, where the early evolution of CMEs can be studied spectroscopically.

In this paper, we investigate the early evolution of a fast CME observed on 2024 August 5 using Fe\,{\sc xiv} 5303~\AA\ spectroscopy from VELC together with AIA  observations. 
We use the VELC spectra to examine the Doppler-shifted emission associated with the erupting plasma and combine these measurements with the POS evolution seen in EUV images.
In particular, we investigate whether the lateral expansion of the CME is uniform in different directions and how its line-of-sight velocity structure is related to the observed POS evolution. 
We further use a simplified three-dimensional model to examine how the CME orientation and non-uniform lateral expansion contribute to the observed spectral structure.


\section{Observations and Data Reduction} \label{sec:style}

\begin{figure*}[t]
    \centering
    \includegraphics[width=\textwidth]{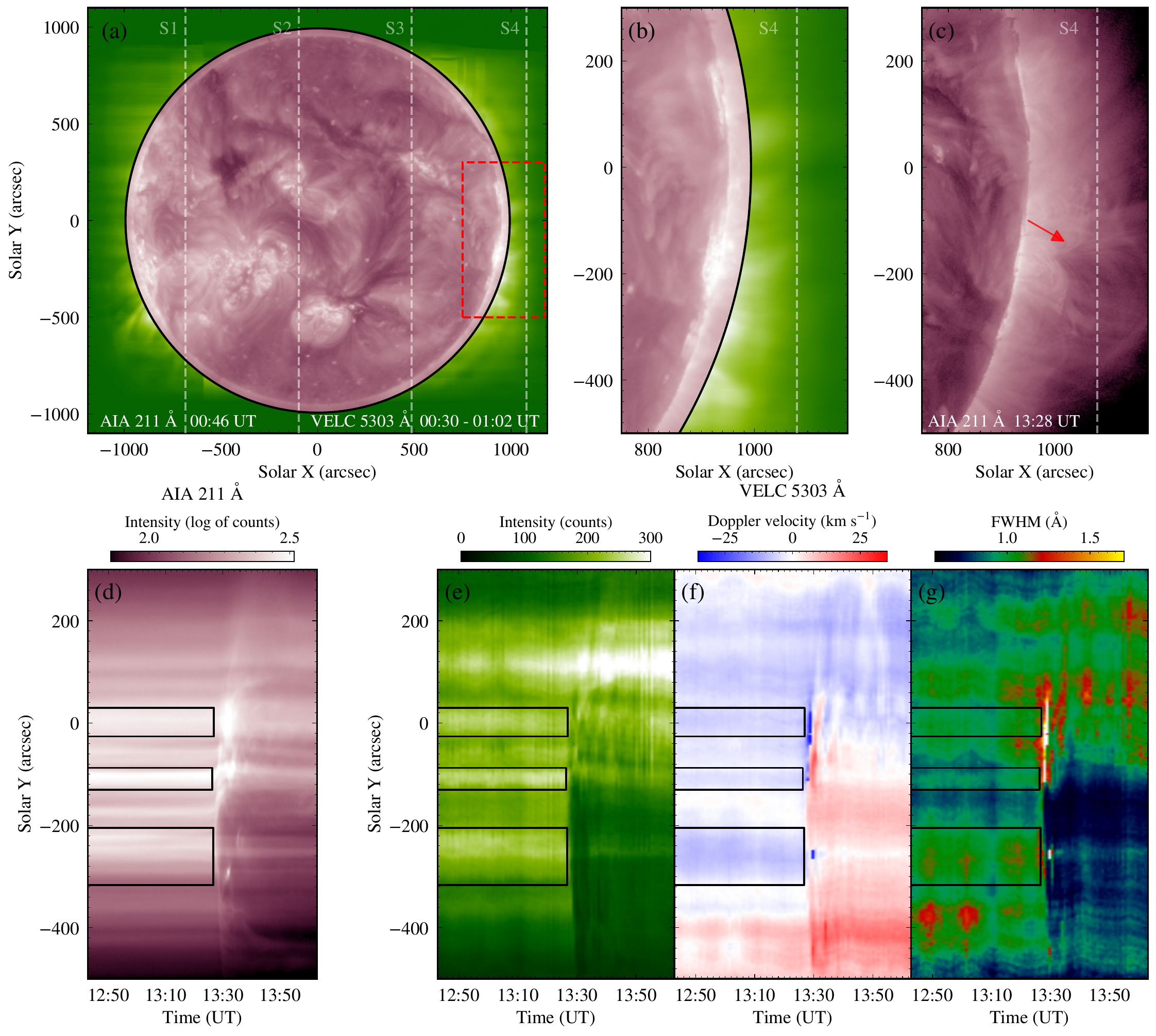}
\caption{VELC and AIA observations of the CME observed on 5~August~2024 and its space--time
evolution at VELC slit S4. (a)~Full-disk composite image
before the eruption: the VELC Fe\,{\sc xiv} 5303~\AA\ raster covering
$1.05$--$1.30~R_\odot$ and 
the spatially binned AIA 211~\AA\ image covering $0$--$1.05~R_\odot$. The four VELC slit positions
are overlaid and the red dashed box marks the western-limb region. (b)~Expanded view of this region. (c)~AIA 211~\AA\ image during the CME eruption, with VELC Slit~S4 (Solar $X \approx 1080^{\prime\prime}$) marked.
(d--g)~Space--time maps along Slit~S4 from
12:42 to 14:03~UT. (d)~AIA 211~\AA\ intensity; (e)~VELC
Fe\,{\sc xiv} 5303~\AA\ peak intensity; (f)~Doppler velocity;
and (g)~line width (FWHM), all derived
from a single-Gaussian fit to the VELC spectra. Examples of the
Gaussian fits are shown in Figure~\ref{fig:fig2}. Three black rectangular boxes highlight the prominent pre-eruption features in the VELC intensity space–time map (panel e) and are overlaid on panels (d)–(g) for easier comparison.}
\label{fig:fig1}
\end{figure*}

On 2024 August 5, active region NOAA 13767 produced an X1.7-class flare at approximately 13:27 UT near the west limb (97$^\circ$W, 8$^\circ$S), followed by a fast CME. The CME was observed by VELC onboard ADITYA-L1, together with AIA.
During this period, the VELC spectrograph operated in a sit-and-stare mode. 
Slit S4 was positioned near the western limb at Solar X $\approx1080\arcsec$ and captured the CME as it propagated through the VELC field of view (Fig.~\ref{fig:fig1}). 
The Fe {\sc xiv} 5303~\AA\ observations have a cadence of 51 s and a spatial sampling of 2.5 arcsec pixel$^{-1}$ along the slit (corresponding to 2-pixel on-board binning), and a spectral sampling of 0.028~\AA\ pixel$^{-1}$. Each slit has a width of 9.6 arcsec.
We use AIA 211~\AA\ images, sensitive to plasma at approximately 2 MK, with a spatial sampling of 0.6 arcsec pixel$^{-1}$ and a cadence of 12 s. A routine VELC raster scan obtained approximately 13 hr before the eruption provides the pre-eruption coronal context (Fig.~\ref{fig:fig1}).

\begin{figure*}[t]
    \centering \includegraphics[width=.95\textwidth,height=0.95\textheight,keepaspectratio]{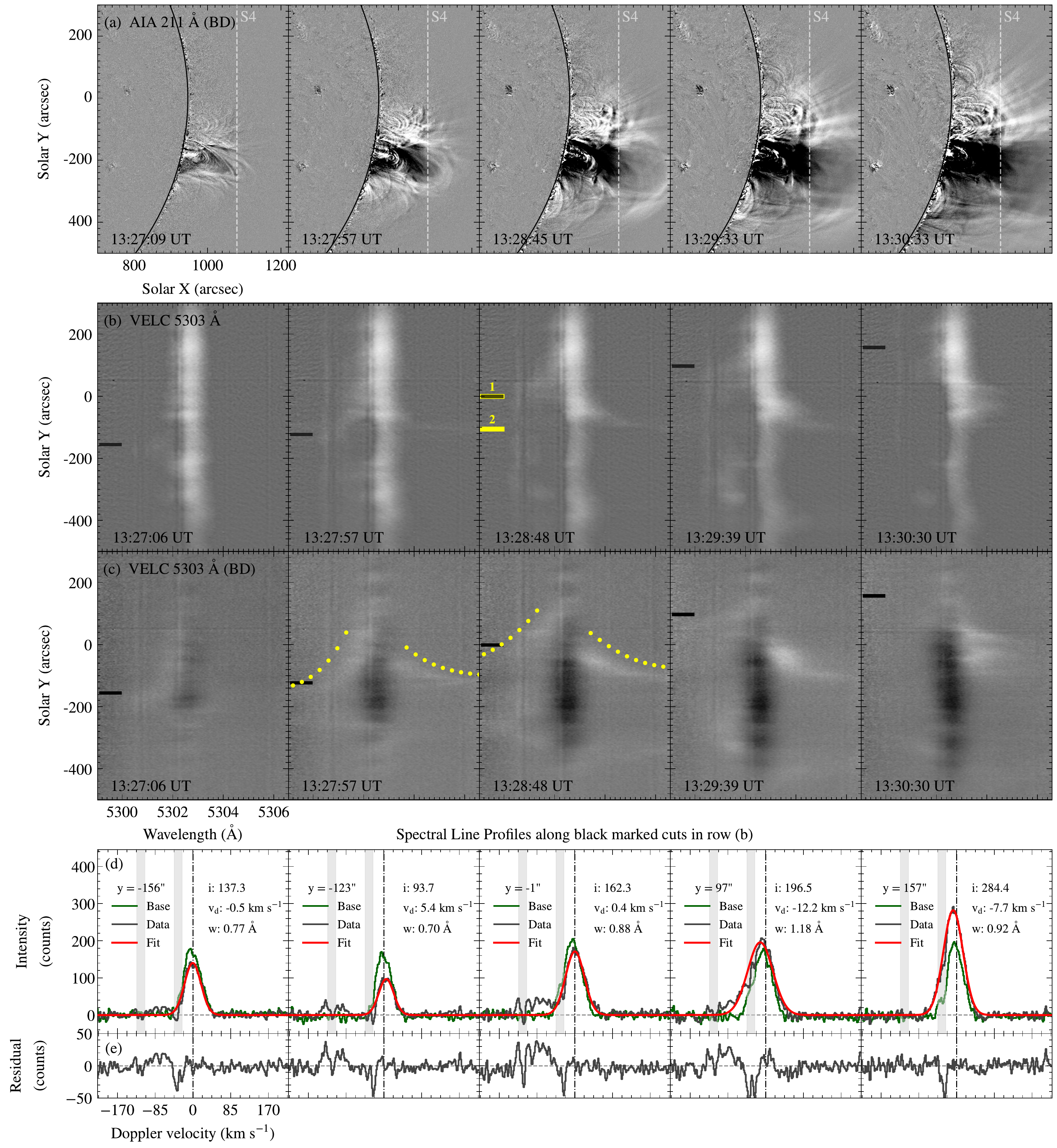}
\caption{Imaging and spectroscopic evolution of the CME along VELC Slit~S4
from 13:27 to 13:30~UT. (a)~AIA 211~\AA\ base-difference images
showing the propagating CME front and the developing coronal
dimming. (b)~VELC Fe\,{\sc xiv} 5303~\AA\ spectral maps along
Slit~S4, showing the emission as a function of wavelength and
solar Y. (c)~Corresponding base-difference spectral maps, which more clearly
highlight the CME-associated intensity depletion near the
Fe\,{\sc xiv} rest wavelength and the bifurcated red- and
blue-shifted structures highlighted by the yellow dotted lines. (d) Representative 
spectra extracted at the spatial locations marked by the black lines
in panel~(b), showing the observed profiles (dark gray), pre-event
reference profiles (green), and single-Gaussian fits to the observed
profiles (red). The fitted intensity ($i$), Doppler velocity ($v_{\rm d}$), and FWHM ($w$) are
also shown. The shaded wavelength regions are excluded from the
fits. (e)~Residuals of the single-Gaussian fits to the observed
profiles. The profiles marked by the yellow line segments in panel~(b) are used for the multi-Gaussian analysis in Figure~\ref{fig:fig3}. An animation showing the time evolution of panels~(a)--(c) of this figure is available online.}
\label{fig:fig2}
\end{figure*}

\begin{figure*}[t]
    \centering
    \includegraphics[width=0.85\textwidth]{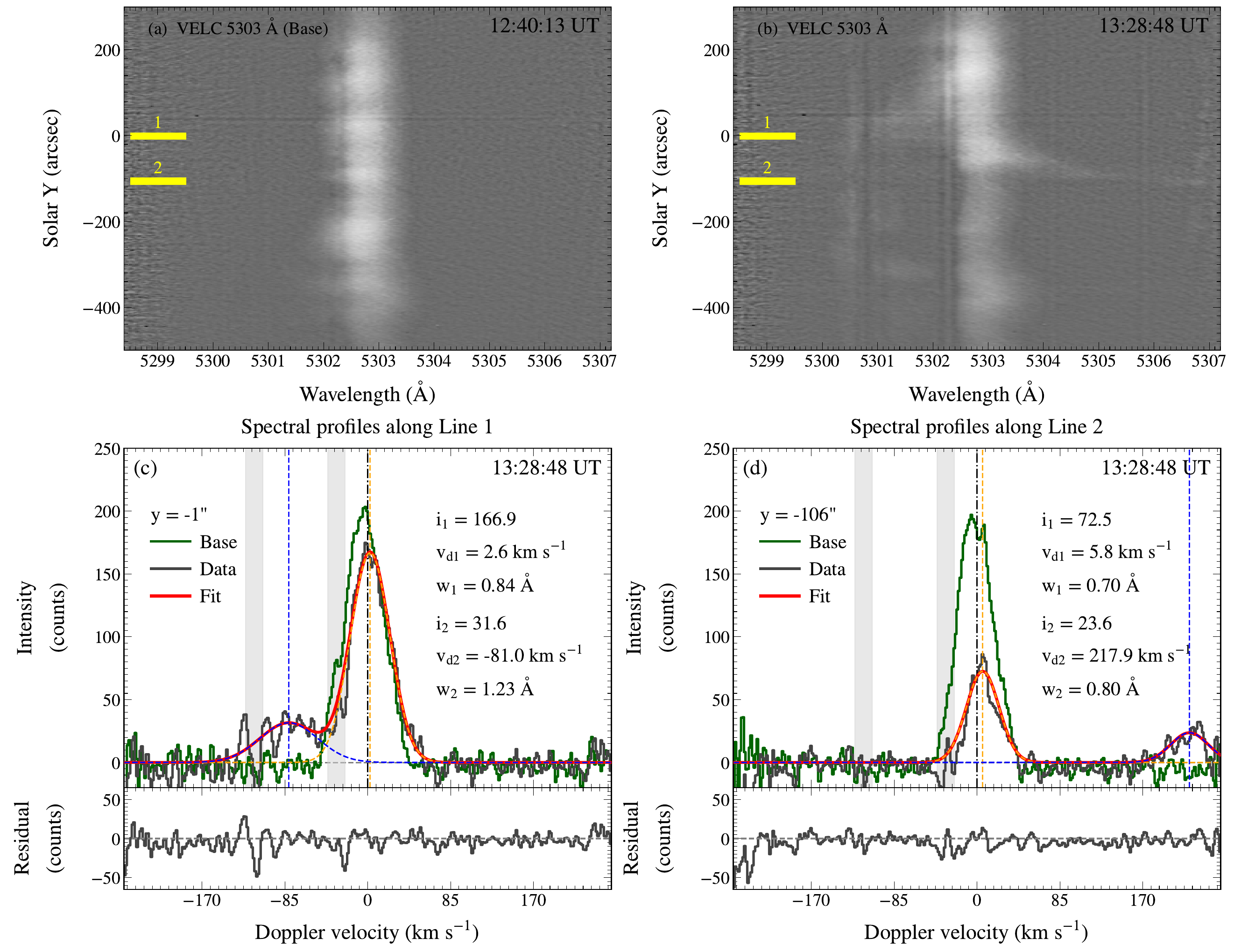}
\caption{Multi-component Gaussian analysis of the VELC
Fe\,{\sc xiv} 5303~\AA\ spectra during the CME passage. (a) and
(b) show spectra along Slit~S4 at the pre-event reference
time (12:40:13~UT) and during the CME passage (13:28:48~UT),
respectively. (c) and (d) present at Solar 
$Y=-1''$ and Solar $Y=-106''$, respectively, at the locations marked by
the yellow line segments in panels~(a) and (b). The observed profiles
(dark gray) are compared with the pre-event reference profiles
(green). The double-Gaussian fits (red) consist of a primary
core component (yellow) and a Doppler-shifted secondary
component (blue). The fitted intensity, Doppler
velocity, and FWHM are printed for each component, with subscripts
1 and 2 denote the primary and secondary components, respectively. The
shaded wavelength regions are excluded from the fits. The lower
panels show the corresponding post-fit residuals.}
\label{fig:fig3}
\end{figure*}

We use Level-1 Fe\,{\sc xiv} 5303~\AA\ spectroscopic data obtained from
the Indian Space Science Data Centre (ISSDC) \citep{2024AdSpR..74..547P,2025SoPh..300...66S}. The data were first
corrected for dark current and detector bias. Since dedicated in-flight
flat-field observations are not available, flat-field and pre-filter
corrections were derived for each slit by averaging VELC raster
observations obtained between July and September 2024.
Wavelength calibration was performed using photospheric absorption lines
from solar disk light scattered off the primary mirror, assuming zero
Doppler shift for these lines. Spectral curvature along the slit was
also corrected. The photospheric contribution was removed using
three-month averaged observations. Small temporal variations in the
photospheric absorption-line depths leave residuals in the processed
spectra. To minimize their influence,
regions spanning approximately 10 pixels around the centers of the two
strong absorption lines were not included in the spectral fitting. These masked regions are shown
by the gray shaded areas in Figures~\ref{fig:fig2}(d) and
\ref{fig:fig3}(c--d). The resulting coronal Fe\,{\sc xiv} emission
profiles were fitted with a single Gaussian to obtain the peak
intensity, Doppler velocity, and line width, using a rest wavelength of
5302.8~\AA\ (Fig.~\ref{fig:fig1}e--g).
Finally, small spatial shifts along the slit caused by satellite drift
were corrected using a cross-correlation technique. Drift perpendicular
to the slit cannot be corrected and causes slow intensity variations
with time in the space--time maps, as the slit samples slightly
different coronal regions during the observing sequence. Some of the
gradual brightening and fading patterns seen in the intensity maps may
therefore arise from this effect (see Fig.~\ref{fig:fig1}e).

\section{Results} \label{sec:style}

\subsection{CME Evolution in AIA and VELC Observations}

Figure~\ref{fig:fig1}(d--g) shows the evolution of the CME along
Slit~S4 from 12:42 to 14:03~UT. The AIA 211~\AA\ space--time map
(Fig.~\ref{fig:fig1}d) shows the onset of the eruption around
13:27~UT, followed by an outward-propagating bright front and a
pronounced intensity depletion between Solar $Y\approx-450''$ and
$50''$. The corresponding decrease in VELC Fe\,{\sc xiv} 5303~\AA\
intensity (Fig.~\ref{fig:fig1}e) indicates depletion of coronal
emission along the slit. As the CME expands, enhanced emission
develops near the propagating front, while depletion and redistribution
of coronal plasma occur in its wake, producing the associated coronal
dimming. The Doppler map shows red- and blue-shifted emission during
the CME passage (Fig.~\ref{fig:fig1}f). The
Fe\,{\sc xiv} line width also increases during the passage of the CME front, with FWHM values exceeding 1.5~\AA\ (Fig.~\ref{fig:fig1}g).

Figure~\ref{fig:fig2} provides a closer view of the CME evolution
along Slit~S4 between 13:27 and 13:30~UT. The AIA 211~\AA\
base-difference images in Figure~\ref{fig:fig2}a show the outward
propagation of the CME-associated front and the progressive
development of the dimming region. The corresponding VELC
Fe\,{\sc xiv} 5303~\AA\ spectral maps in Figure~\ref{fig:fig2}b
show the evolution of the emission simultaneously in 
the wavelength--height plane. At the beginning of the sequence, the emission is
concentrated close to the Fe\,{\sc xiv} rest wavelength. As the CME
front crosses the slit, enhanced emission develops over an extended
range of Solar $Y$ and wavelength, and the spectral structure changes
with time as the CME propagates.

The CME-associated spectral changes are more clearly seen in the
base-difference maps shown in Figure~\ref{fig:fig2}c. A pronounced
reduction in the spectral intensity relative to the pre-eruption
state is seen during the CME passage, particularly around the
strong coronal emission near the Fe\,{\sc xiv} 5303~\AA\ line. 
Superposed on this depletion, transient emission extends into both
the red and blue wings. Around 13:27:57~UT, the wing emission develops
into two distinct branches on opposite sides of the Fe\,{\sc xiv}
rest wavelength. The branches are also separated along the
Solar $Y$ direction, with the red-shifted emission predominantly
appearing at lower heights and the blue-shifted emission at higher
heights. They have different spectral extents, and their positions
and shapes evolve as the CME propagates. The evolving red- and
blue-shifted branches, highlighted by the yellow dotted curves in
Figure~\ref{fig:fig2}(c), form the characteristic two-branch spectral
structure. We refer to this evolving structure as the bifurcated
spectral morphology.

The one-dimensional profiles extracted at selected locations within these structures (Figure~\ref{fig:fig2}d) provide a closer view of the spectral features seen in Figure~\ref{fig:fig2}(c). The profiles were spatially averaged over three adjacent slit pixels
and smoothed using a three-point running average in wavelength before
fitting. The selected profiles show
additional wing emission compared with the pre-eruption profiles,
with its direction and extent varying with Solar $Y$. The
single-Gaussian fits reproduce the main coronal emission near the
line center but leave systematic residuals in the wings
(Fig.~\ref{fig:fig2}e), indicating additional spectral components
associated with the CME. We therefore examine these kinds of profiles using
multi-component Gaussian fitting in the following section.

\subsection{Multi-component Spectral Fitting}

Figure~\ref{fig:fig3} shows the Fe\,{\sc xiv} 5303~\AA\ profiles at
two representative locations along Slit~S4 during the CME passage
(13:28:48~UT), together with the corresponding pre-eruption profiles.
Each profile was fitted with a linear background, a primary
Gaussian component representing the ambient coronal emission, and an
additional Gaussian component to account for the Doppler-shifted
CME-associated emission. 

At Solar $Y \approx -1''$ (Fig.~\ref{fig:fig3}c), the primary component has a
Doppler velocity of $+2.6~\mathrm{km\,~s^{-1}}$ and FWHM of
$0.84$~\AA\, while the secondary component is blueshifted by
$-81.0~\mathrm{km\,~s^{-1}}$ with FWHM of $1.23$~\AA. At
Solar $Y \approx -106''$ (Fig.~\ref{fig:fig3}d), the primary component has
$+5.8~\mathrm{km\,~s^{-1}}$ and FWHM of $0.70$~\AA\,
while the secondary component is redshifted by
$+217.9~\mathrm{km\,~s^{-1}}$ with FWHM of $0.80$~\AA\. The primary components remain close to the rest wavelength, while the secondary components are strongly Doppler shifted and have broader line widths. Their spatial separation is consistent with
the two-branch structure identified in Figure~\ref{fig:fig2}.

\begin{figure*}[t]
    \centering
    \centering \includegraphics[width=\textwidth,height=0.8\textheight,keepaspectratio]{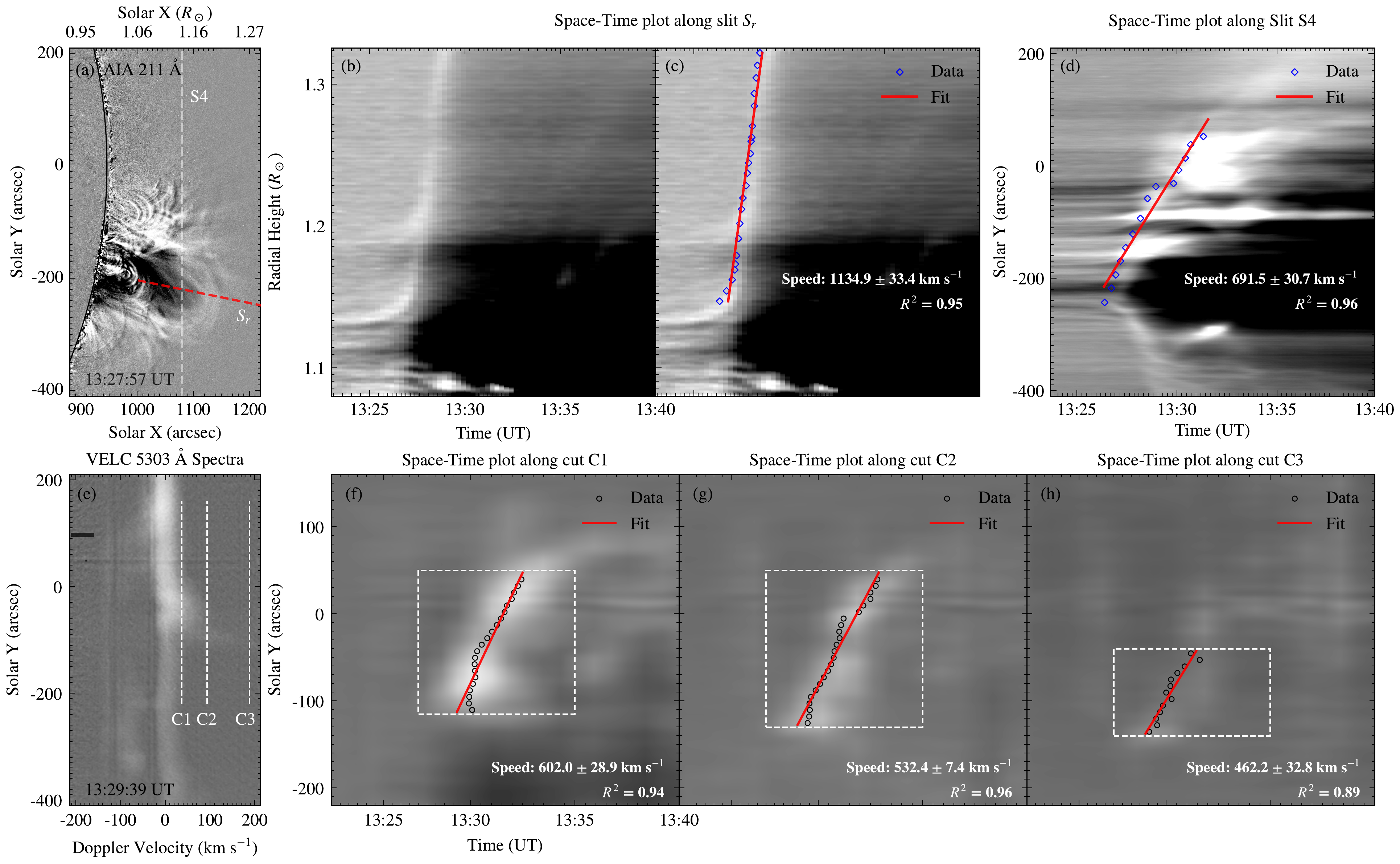}
\caption{Radial and lateral propagation of the CME and
Doppler-resolved propagation of VELC emission features. (a)~AIA
211~\AA\ image showing the artificial slit (red dashed line, $S_r$)
placed along the CME radial propagation direction, together with VELC
Slit~S4 (white dashed line). (b)~AIA 211~\AA\ base-difference
space--time map along the artificial slit $S_r$. (c)~Same as (b), with
the manually tracked CME propagating front (blue diamonds) and its
linear fit (red line). (d)~AIA 211~\AA\ base-difference space--time
map along VELC Slit~S4, showing the manually tracked lateral front
(blue diamonds) and its linear fit (red line). 
(e)~VELC Fe\,{\sc xiv} 5303~\AA\
spectrum at 13:29:39~UT, with three artificial cuts (C1--C3) marked
at Doppler velocities of 37, 93, and 192~km\,s$^{-1}$, respectively. 
 (f--h)~Space--time intensity maps
constructed from the three cuts. For cuts C1 and C2, the emission is averaged over a $\pm14~\mathrm{km\,~s^{-1}}$ Doppler-shift range around their central velocities, while for cut C3 it is averaged over a $\pm28~\mathrm{km\,~s^{-1}}$ range.
The positions of the emission peaks (black circles) 
are obtained from Gaussian fits to the intensity profiles at each
height within the marked regions (white
dashed rectangles), and the resulting tracks are fitted
linearly (red lines) to determine the propagation speeds. The measured propagation speeds and corresponding
$R^2$ values are indicated in the respective panels.}
\label{fig:fig4}
\end{figure*}
\begin{figure*}[t]
    \centering
    \includegraphics[width=\textwidth]{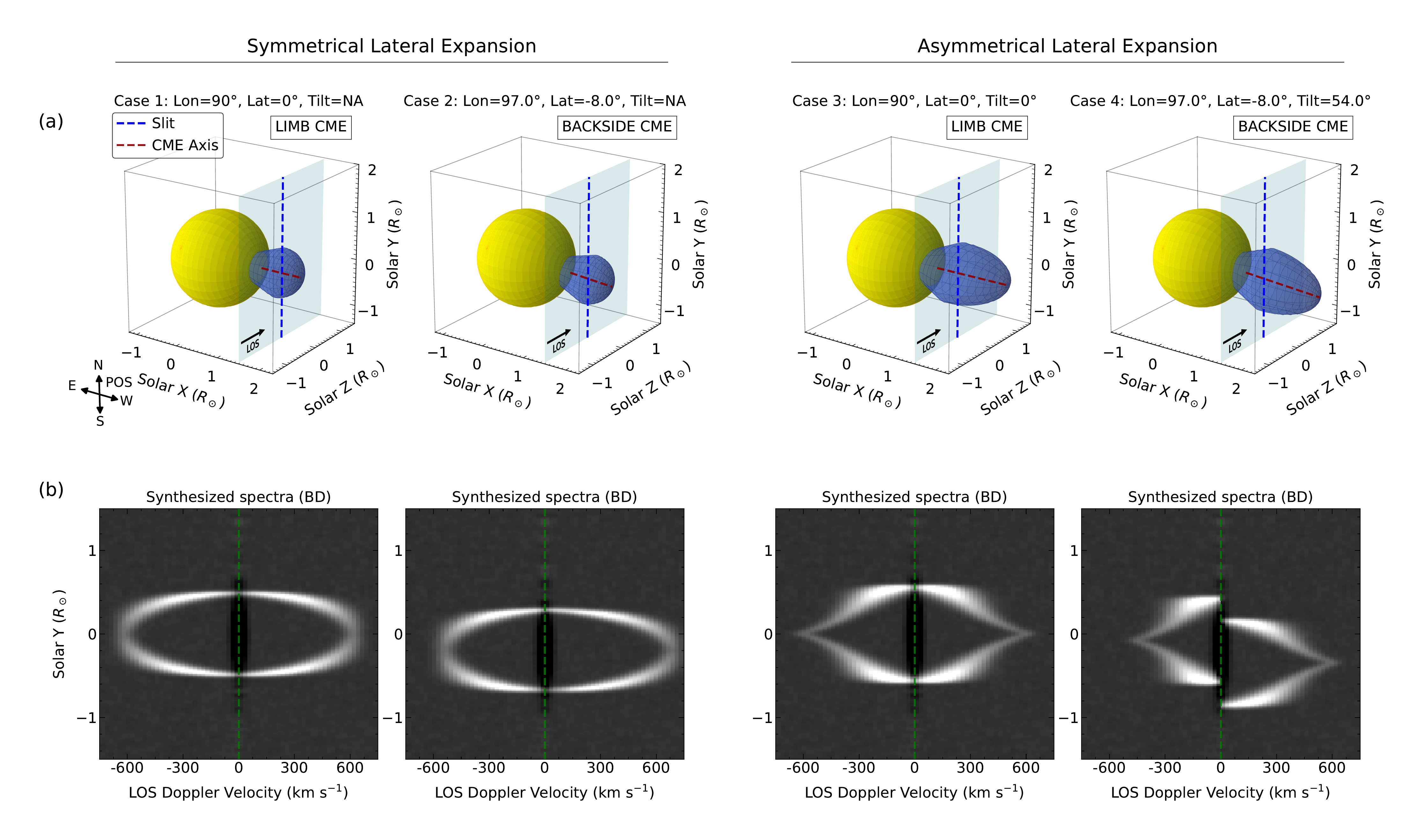}
\caption{Three-dimensional modeling of the CME geometry and
synthesized spectra for four model configurations. The upper
panels (a) show the CME geometry, with the yellow sphere representing
the Sun and the blue structure representing the CME. The red dashed
lines indicate the CME axis. The line-of-sight (LOS) emission-integration
plane is highlighted by the shaded plane, which also shows the slit
orientation marked in blue. Cases~1 and~2 assume symmetric lateral
expansion, while Cases~3 and~4 include asymmetric lateral expansion.
Cases~1 and~3 represent a limb CME, whereas Cases~2 and~4 represent a
backside CME with the GCS-derived CME orientation. The GCS-derived
orientation corresponds to a CME located $7^\circ$ behind the solar
limb, is referred to as the ``Backside CME'' configuration. The
lower panels (b) show the synthesized base-difference (BD) spectral maps along the slit, as a function of
LOS Doppler velocity and projected Solar $Y$ position. The LOS Doppler velocity and intensity in the synthetic spectra are shown in arbitrary scales. The results illustrate how CME orientation, axis tilt and lateral
expansion modify the spatial and spectral morphology of the emission. An animation of this figure showing the synthetic spectral evolution for all four
configurations is available online.}
    \label{fig:fig5}
\end{figure*}

\subsection{Radial and Lateral Propagation}
The LOS velocities of the erupting plasma were established from the
spectral analysis above. We next examine the radial and lateral
propagation of the CME using the complementary AIA and VELC
observations. Figure~\ref{fig:fig4} summarizes these measurements.

Figure~\ref{fig:fig4}(a--c) shows the radial propagation of the CME in
the POS. An artificial slit at a position angle of
$101.5^\circ$ was placed along the propagation direction
(Fig.~\ref{fig:fig4}a), and the CME leading edge was tracked in the
corresponding space--time map (Fig.~\ref{fig:fig4}b--c). A linear
fit to the manually tracked bright propagating front gives a projected
radial propagation speed of $1134.9 \pm 33.4~\mathrm{km\,~s^{-1}}$.

To examine the lateral evolution, we use the co-spatial AIA and VELC
observations along Slit~S4. Tracking the propagating front in the AIA
space--time map gives a projected POS propagation speed of
$691.5 \pm 30.7~\mathrm{km\,~s^{-1}}$.

The VELC spectra further allow the propagation of emission features to
be examined over selected Doppler-velocity ranges.  We define three artificial slits (C1--C3) in the red wing of the Fe\,{\sc xiv} profile, corresponding to Doppler velocities of $+37$, $+93$, and $+192~\mathrm{km\,~s^{-1}}$, respectively, to investigate the spatiotemporal evolution of the CME-associated emission. The red wing is used because these
features have higher signal-to-noise ratios and more coherent ridges
than the corresponding blue-wing features. The ridge positions were
obtained by fitting the intensity profile at each height with a
Gaussian, followed by linear fits to the resulting height--time
tracks. Although a blue-shifted component is clearly visible in the
individual spectra, the corresponding blue-wing space--time features
are too diffuse for a reliable propagation-speed measurement. 

The three red-wing velocity ranges give projected POS
propagation speeds of $\sim$602, 532, and
462$~\mathrm{km\,~s^{-1}}$, respectively. The projected
propagation speed therefore decreases as the redshifted velocity
increases. This trend does not necessarily imply a decrease in the
actual plasma speed, but can result from the changing direction of
the expansion velocity across the CME shell. Since Slit~S4 samples
the expanding CME shell at a fixed location, different positions
along the slit intersect different parts of the shell. Near the
POS-lateral front of the expanding shell sampled along the slit, the
expansion velocity is predominantly in the POS, giving a larger
projected POS speed and a smaller LOS component. Toward the central
part of the shell sampled by the slit, the expansion velocity has a
larger LOS component, producing a larger Doppler shift and a smaller
projected POS speed. The velocity-resolved propagation therefore
indicates a systematic change in the direction of the expansion
velocity across the observed CME shell.

This behavior is consistent
with an expanding CME in which the lateral velocity has components
both in the POS and along the LOS, as explored with the
three-dimensional model in Section~\ref{sec:model_text}.


\subsection{Three-dimensional Modeling of the CME}
\label{sec:model_text}
To investigate the geometric origin of the bifurcated VELC spectra, we
use a parameterized three-dimensional kinematic model of an expanding
CME. The model consists of a shell-like structure undergoing radial
propagation and lateral expansion, with the lateral velocity specified
independently along two perpendicular directions.
The model geometry and velocity prescription are described in Appendix~\ref{app:model}.

We synthesize the Fe\,{\sc xiv} 5303~\AA\ emission by integrating the
optically thin emission along the line of sight. The synthetic spectra
are used only for a qualitative comparison with the VELC observations.
The model parameters are representative and are not intended to provide
a quantitative or unique reconstruction of the three-dimensional CME
geometry or velocity field.

Figure~\ref{fig:fig5} shows four representative configurations designed
to separate the effects of CME orientation and lateral expansion.
Case~1 represents a limb CME with symmetric lateral
expansion, producing approximately symmetric red- and blue-shifted
components with a semicircular morphology. Case~2 introduces the
heliographic location obtained from the GCS reconstruction (Appendix~\ref{app:gcs})
while retaining symmetric lateral expansion. The resulting projection
produces unequal Doppler shifts. For Cases~1 and 2, the lateral cross-section is circular because the expansion is symmetric, so the axis tilt is not applicable. Case~3 retains the limb geometry but introduces asymmetric
lateral expansion, with a larger expansion speed along the LOS than
along the Solar~Y direction. This changes the spectral shape, producing
a more pointed morphology. Case~4 combines the GCS-constrained
orientation, including axis tilt, with asymmetric lateral expansion and produces a spectral
pattern most similar to that observed by VELC. 

The comparison shows that the CME orientation, axis tilt and lateral expansion
affect different aspects of the observed spectral structure. The axis tilt and orientation primarily determines the relative displacement and Doppler
asymmetry of the red- and blue-shifted components, respectively, whereas the
distribution of lateral expansion controls their spectral extent and
shape. The accompanying animation shows the synthetic spectral
evolution for all four configurations. Among them, Case~4 also shows a
spatiotemporal evolution of the bifurcated spectral structure that
closely resembles the VELC observations. Thus, the observed spectral
morphology and its evolution are consistent with an inclined axially tilted CME
undergoing non-uniform lateral expansion. 


\section{Discussion}
\label{sec:discussion}
The AIA and VELC observations provide complementary constraints on
the CME evolution in the low corona. AIA 211~\AA\ traces the
projected propagation of the CME front and the associated coronal
dimming, while VELC spectroscopy reveals the Doppler-shifted plasma
motions as the CME crosses Slit~S4. The combination therefore provides
information on both the POS propagation and the LOS velocity structure
of the expanding CME. 

During the CME passage, the VELC Fe\,{\sc xiv} profiles develop
distinct red- and blue-shifted components at different locations
along Slit~S4. Their spatial separation indicates that different portions of the
erupting structure contribute to the observed Doppler emission,
revealing multiple velocity components within the expanding CME. The finite spectral range limits the LOS velocity
measurement: the Doppler-shifted secondary component reaches the edge of
the available spectral window, so its fitted centroid, corresponding
to a Doppler shift of {$+217.9~\mathrm{km\,~s^{-1}}$} is likely a lower
limit to the LOS velocity of the receding plasma. Despite this
limitation, the distinct bifurcated spectral shape and its
spatiotemporal evolution provide important information on the LOS
component of the CME lateral expansion.

The combined AIA and VELC observations show that the CME evolves
non-uniformly in the low corona. The projected radial propagation speed
of the CME front is $1134.9~\mathrm{km\,~s^{-1}}$, compared with
a projected POS propagation speed of
$691.5~\mathrm{km\,~s^{-1}}$ along Slit~S4. 
This difference indicates that the CME expansion is not uniform in
different directions, consistent with previous observations of
anisotropic and non-self-similar CME expansion in the low corona
\citep{2020A&A...635A.100C,2020ApJ...899....6M}.

The Doppler-selected
red-wing features propagate across the slit at $\sim 602$ -- $462~\mathrm{km\,~s^{-1}}$, with the projected propagation speed
decreasing as the redshifted velocity increases. This trend does not
necessarily imply that plasma with a larger LOS velocity is intrinsically
slower. Rather, the Doppler-selected features can sample different
parts of the expanding CME shell, where the velocity vector has
different projections onto the LOS and POS. Plasma near the POS-lateral front of the expanding shell, as sampled along the slit, has a larger POS component and a smaller LOS component, whereas plasma closer to the center of the expanding structure has a larger LOS component and a smaller projected POS component. The
velocity-dependent propagation therefore indicates a change in the
direction of the velocity vector across the expanding CME.

The three-dimensional model provides a geometric interpretation
of the observed spectral structure. A model with symmetric lateral expansion does not reproduce the observed spectral morphology as closely as a model with different expansion rates along two perpendicular directions. The CME orientation affects the LOS contribution from the
radial and axis tilt affects lateral motions and the displacement of the red- and
blue-shifted components, while asymmetric lateral expansion changes
their spatial and spectral extent. The combination of the observed CME
orientation, axis tilt and asymmetric lateral expansion produces a spectral
structure similar to that observed by VELC. The accompanying animation
shows the synthetic spectral evolution for all four model
configurations. Case~4, which combines the observed CME orientation (GCS-derived) with asymmetric lateral expansion, produces a bifurcated spectral morphology that shows a close visual resemblance to the VELC observations in both spatial structure and temporal evolution.

Note that the lower branch of the observed bifurcated spectral structure is relatively faint and cannot be clearly distinguished throughout the event because of its low signal-to-noise ratio. However, a closer inspection of the animation (Fig.~\ref{fig:fig2}) reveals faint 
signatures of this lower branch and suggests an asymmetric spectral extent, with the right branches extending farther than the left. The two Doppler-shifted branches on the right also appear slightly displaced toward lower heights relative to their counterparts on the left. These features show a close visual resemblance to the spectral evolution produced by Case~4. Furthermore, the comparison is limited by the finite spectral range of the VELC observations. The LOS velocity in Figure~\ref{fig:fig5}(b) is shown on an arbitrary scale, with synthetic spectra extended to $\sim\pm750~\mathrm{km\,~s^{-1}}$, to illustrate the bifurcated spectral structure over a broader velocity range than VELC observes ($\sim\pm235~\mathrm{km\,~s^{-1}}$; Fig.~\ref{fig:fig2}c). The pointed features at the edges of the synthetic spectra therefore likely reflect emission that VELC does not detect in the late evolution phase. Nevertheless, the model spectra and the observed spectral evolution show close agreement in their spatiotemporal evolution within the inner spectral range.

One possible explanation for the non-uniform lateral expansion is an
asymmetric magnetic environment around the erupting CME. The expanding
structure is confined by the magnetic pressure and tension of the
surrounding corona, and an imbalance in these forces can allow greater
expansion toward regions of weaker magnetic confinement. Such an
asymmetric magnetic environment has been suggested as a possible cause
of non-uniform CME expansion \citep{2011LRSP....8....1C}. Recent
observations also show lateral deformation of CME structures
associated with their interaction with the surrounding magnetic field
\citep{2026ApJ...997..303H}. The asymmetric expansion inferred in this
event may therefore be related to non-uniform magnetic confinement in
the surrounding corona, although the present observations do not
provide sufficient constraints to establish this connection.


\section{Summary}
\label{sec:highlight}

We investigated the early evolution of a fast CME observed on 2024 August 5
using Fe\,{\sc xiv} 5303~\AA\ spectroscopy from VELC together with
AIA observations. AIA provides the projected radial and lateral propagation speeds in the POS, while VELC provides the LOS
velocity structure of the expanding plasma. The VELC spectra reveal
Doppler-shifted emission associated with the CME leading edge, with
oppositely directed LOS motions producing a bifurcated spectral
structure. The Doppler-resolved propagation and spatial variation of
the spectral components indicate that the velocity direction changes across
the expanding CME. Although the Doppler-shifted component reaches the
edge of the available spectral range and its maximum LOS velocity is
likely not captured, its spectral shape and spatial evolution provide
information on the LOS component of the CME lateral expansion. A three-dimensional model shows that an inclined CME  with axis tilt and asymmetric lateral expansion can produce a spectral structure similar
to that observed. Together, the observations and modeling support a picture of
non-uniform three-dimensional CME expansion, with different radial,
POS-lateral, and LOS-lateral velocity components.

\begin{acknowledgments}
We acknowledge the VELC payload team at the Indian Institute of Astrophysics (IIA) and ISRO for the successful development, operation of the \textit{ADITYA-L1} mission, and for providing the data. Aditya-L1 is an observatory class mission which is fully funded and operated by the Indian Space Research Organization (ISRO). The mission was conceived and realised with the help from various ISRO centres. The science payloads and science ready data products are realised by the payload PI institutes in close collaboration with ISRO centres. The PI institutes are: Indian Institute of Astrophysics (IIA); Inter University Centre for Astronomy and Astrophysics (IUCAA), Laboratory for Electro-optics Systems (LEOS/URSC); Physical Research Laboratory (PRL); U R Rao Satellite Centre (URSC); and Space Physics Laboratory (SPL/VSSC). We acknowledge the use of data from the Aditya-L1 mission of the Indian Space Research Organisation (ISRO), archived at the Indian Space Science Data Centre (ISSDC).  We also acknowledge the use of extreme-ultraviolet imaging data from the Solar Dynamics Observatory (SDO), courtesy of NASA and the AIA science team. The SOHO/LASCO data used in this study are produced by a consortium of the Naval Research Laboratory (USA), Max-Planck-Institut für Sonnensystemforschung (Germany), Laboratoire d'Astrophysique de Marseille (France), and the University of Birmingham (UK); SOHO is a project of international cooperation between ESA and NASA.
\end{acknowledgments}

\appendix
\renewcommand{\thefigure}{A\arabic{figure}}
\setcounter{figure}{0}


\section{Three-dimensional CME Model and Synthetic Spectra}
\label{app:model}

We use a parameterized three-dimensional kinematic model to examine how
the geometry and velocity field of an expanding CME produce the observed
Doppler-shifted spectral structure. The model is intended for a
qualitative comparison with the VELC observations and is not used to
derive the three-dimensional geometry or velocity field of the observed
CME. The adopted parameters are therefore representative.

\subsection{CME Geometry}

\begin{figure*}[t]
\centering
\includegraphics[width=\textwidth]{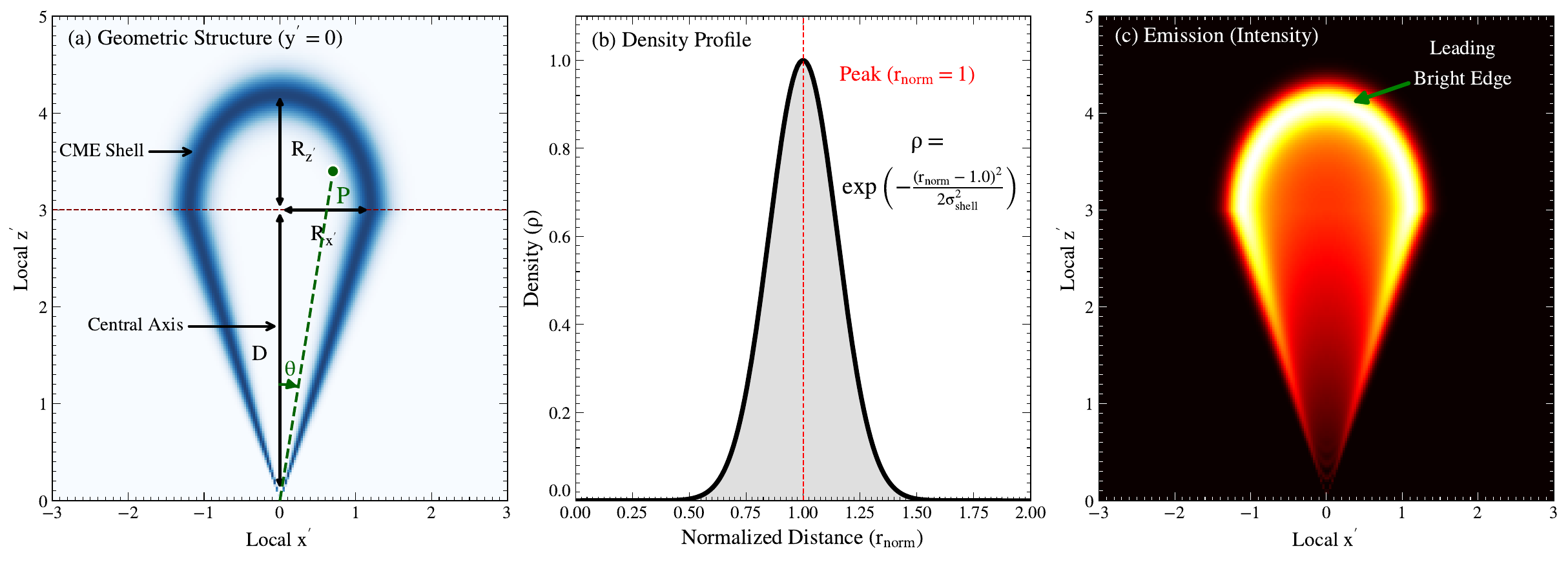}
\caption{Geometry and density distribution of the CME model and the corresponding synthetic emission.
Panel (a) illustrates the 3D cone--hemispherical cross-section at $y'=0$, 
parameterized by the conical leg height ($D$), lateral thickness
($R_{x'}$), apex cap height ($R_{z'}$), and shell angle ($\theta$)
measured from the cone origin relative to the central axis for a shell
point P. Panel (b) shows the density profile $\rho(r_{\rm norm})$ of CME shell
as a function of the normalized distance
$r_{\rm norm}$ from the central axis, with $r_{\rm norm}=1$ representing position of density peak. Panel (c) shows the synthetic emission integrated along the $y'$ direction.}
\label{fig:fig_a1}
\end{figure*}

The CME is represented by a shell-like structure consisting of a conical
section capped by a hemisphere, as illustrated in
Figure~\ref{fig:fig_a1}. The tip of the conical section is anchored at
the center of the Sun, and the uppermost point of the hemispherical cap
defines the CME apex. The modeled structure consists of a high density-shell with finite thickness and a relatively
low-density interior. The density is assumed to peak at a reference
surface and to decrease away from this surface following a Gaussian
profile. Thus, most of the emission
originates from a finite-thickness shell.

We introduce a local Cartesian coordinate system $(x',y',z')$ attached
to the CME. The origin is located at the vertex of the cone, and the $+z'$-axis is directed along the radial propagation
direction toward the CME apex. A point on the CME surface is specified
by the shell angle $\theta$ measured from the $+z'$-axis and the
azimuthal angle $\omega$ around the $z'$-axis. The shell angle varies
from $\theta=0^\circ$ at the apex to $|\theta| < 90^\circ$ toward the
lateral flanks.

\subsection{Velocity Field}

The CME velocity field consists of radial propagation and lateral
expansion. The radial velocity, $v_{\rm rad}$, is directed along the
CME propagation axis, while the lateral velocity, $v_{\rm lat}$, is
perpendicular to this axis. The radial component dominates near the
apex, whereas lateral expansion becomes increasingly important toward
the flanks. We describe the transition between these two components as

\begin{equation}
v(\theta,\omega)
=
v_{\rm lat}(\omega)
+
\left[v_{\rm rad}-v_{\rm lat}(\omega)\right]
e^{-A|\theta|},
\label{eq:velocity_field}
\end{equation}

\noindent
where $A$ controls the transition from radial propagation near the
apex to lateral expansion toward the flanks.

To allow for asymmetric lateral expansion, the lateral velocity is
specified by two orthogonal components, $v_{\rm lat,x'}$ and
$v_{\rm lat,y'}$. The effective lateral expansion velocity at azimuth
$\omega$ is

\begin{equation}
v_{\rm lat}(\omega)
=
\frac{
v_{\rm lat,x'}v_{\rm lat,y'}
}{
\sqrt{
\left(v_{\rm lat,y'}\cos\omega\right)^2+
\left(v_{\rm lat,x'}\sin\omega\right)^2
}
}.
\label{eq:lateral_velocity}
\end{equation}

For $v_{\rm lat,x'}=v_{\rm lat,y'}$, the lateral expansion is
symmetric. Unequal values produce an elliptical cross-section and
asymmetric expansion. The velocity components in the local CME
coordinate system are then

\begin{align}
v_x' &= v(\theta,\omega)\sin\theta\cos\omega, \\
v_y' &= v(\theta,\omega)\sin\theta\sin\omega, \\
v_z' &= v(\theta,\omega)\cos\theta.
\label{eq:velocity_components}
\end{align}

\subsection{CME Orientation}

The local CME coordinate system is transformed to a heliocentric
Cartesian coordinate system $(X,Y,Z)$, where the $Z$-axis is directed
along the observer's LOS and the $XY$-plane defines the POS. The position of the CME on the solar surface is specified by its
heliographic longitude $\phi$ and latitude $\lambda$. These angles
determine the orientation of the CME propagation axis with respect to
the observer.

An additional tilt angle $\gamma$ describes the rotation of the CME
cross-section about the propagation axis. This parameter is relevant only when the cross-section is elliptical, as arises in our model due to non-uniform lateral expansion. The resulting transformation
determines the orientation of the CME with respect to the plane of the
sky and the line of sight. For the configurations representing the observed CME, we adopt
$\phi=97^\circ$, $\lambda=-8^\circ$, and $\gamma=54^\circ$,
based on the GCS reconstruction described in Appendix~\ref{app:gcs}.

\subsection{Synthetic Fe\,{\sc xiv} Spectra}

To compare the model with the VELC observations, we synthesize the
Fe\,{\sc xiv} 5303~\AA\ emission from the model CME. The calculation
includes only the prescribed geometry, density distribution, and velocity field.
We do not model the thermodynamic evolution, ionization state, or
detailed excitation conditions of the CME. Instead, we assume optically
thin emission produced by collisional excitation and use the model
density to prescribe the relative emissivity.

The synthetic spectral intensity at a position $(X,Y)$ is obtained by
integrating the emission along the line of sight,

\begin{equation}
I(X,Y,\nu)
=
\int
\epsilon(X,Y,Z)\,
\mathscr{G}
\left[
\nu-V_{\rm LOS}(X,Y,Z)
\right]
\,dZ,
\label{eq:synthetic_spectrum}
\end{equation}

where $\nu$ is the Doppler velocity used as the spectral coordinate,
$V_{\rm LOS}$ is the local LOS velocity, and $\epsilon$ is
the local emissivity. We assume

\begin{equation}
\epsilon \propto n^2,
\label{eq:emissivity}
\end{equation}

where $n$ is the model density. The function $\mathscr{G}$ represents
a Gaussian local emission profile centered at the local LOS
velocity. The intrinsic width is assumed to be constant throughout
the CME.

Because the density is concentrated in the shell, the synthesized
emission is dominated by regions where the line of sight intersects
the CME shell. The resulting spectra therefore depend on both the
orientation of the shell and the LOS component of its
velocity field.

For comparison with the observations, synthetic spectra at the initial
time are subtracted from those at subsequent times to produce
base-difference (BD) spectra. For constructing synthetic BD spectra, the initial state is taken as the pre-eruption reference state and have no CME-associated bulk motion.

\begin{figure*}[t]
    \centering
    \includegraphics[width=\textwidth,height=0.5\textheight,
    keepaspectratio]{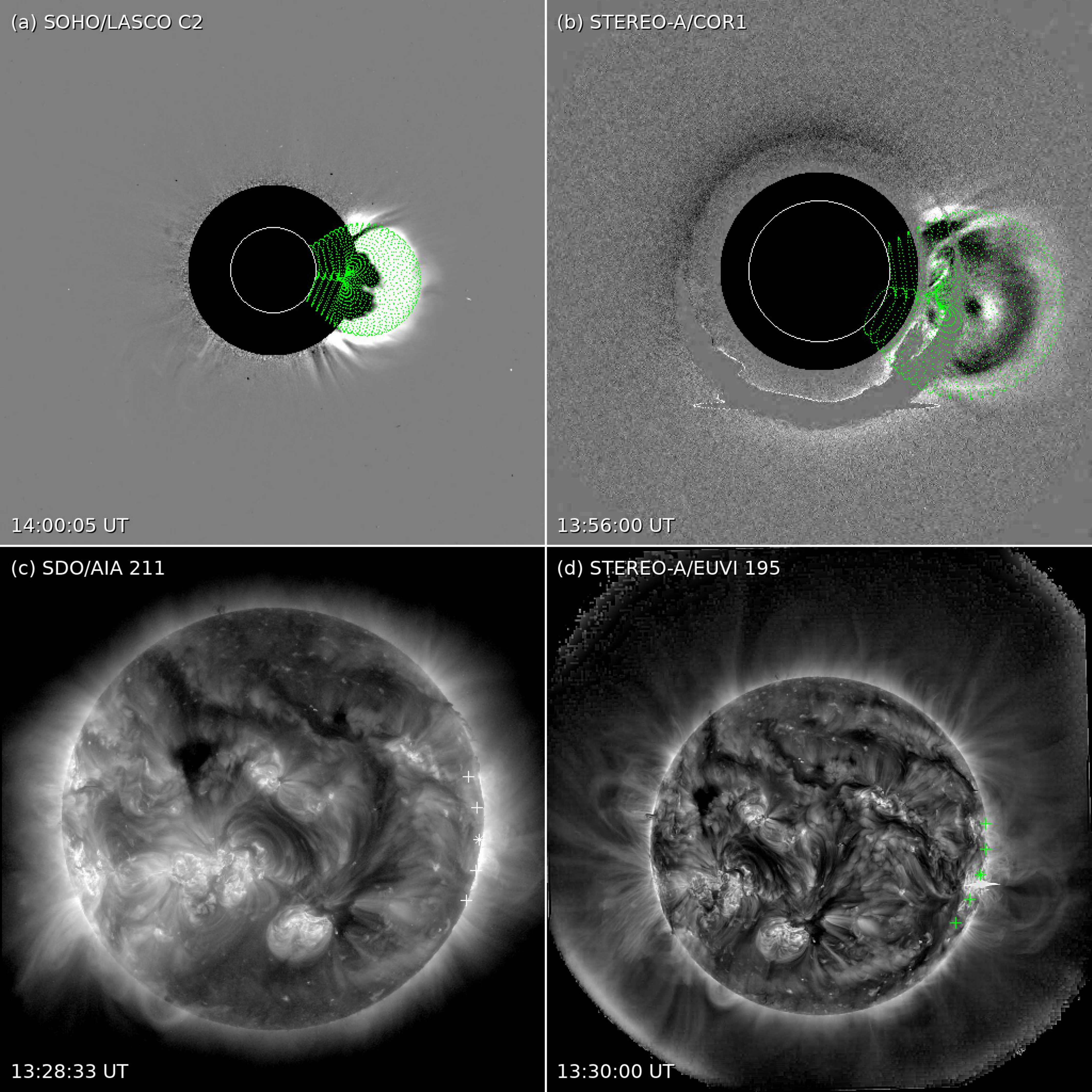}
   \caption{GCS reconstruction of the 5~August~2024 CME using
multi-viewpoint coronagraph and EUV observations. The best-fit GCS
wireframe (green mesh) is overlaid on the observed CME in the
coronagraph panels. The reconstruction gives a CME longitude of
$97^\circ$, latitude of $-8^\circ$, and an axis tilt of $54^\circ$.
The green ``+'' symbols in the lower panels mark the corresponding
footpoint locations. These GCS-derived parameters are used to set the
CME orientation in our three-dimensional CME model.}
    \label{fig:fig_a2}
\end{figure*}
\subsection{Model Configurations}
\label{app:model_config}

We consider four representative configurations to separate the effects
of CME orientation and asymmetric lateral expansion. The corresponding
geometries and synthesized base-difference spectra are shown in
Figure~\ref{fig:fig5}.

The first configuration represents a CME at the solar limb with  symmetric lateral expansion. The front and rear portions of the
expanding shell produce oppositely directed LOS velocities. The
resulting red- and blue-shifted components form an approximately
symmetric, semicircular pattern. The LOS velocity is largest near the
center of the CME, where the lateral expansion has its largest LOS
projection, and decreases toward the upper and lower boundaries where
the expansion becomes increasingly perpendicular to the LOS.

The second configuration uses the heliographic location 
obtained from the GCS reconstruction while retaining symmetric lateral
expansion. The CME is located behind the limb and propagates partly
away from the observer, introducing an additional LOS component from
the radial propagation. As a result, the red- and blue-shifted
components have different Doppler amplitudes.

The third configuration has the same limb location as
the first case, and without any axis tilt, but introduces asymmetric lateral expansion. The lateral
expansion speed along the LOS is larger than that along the Solar~Y
direction. Consequently, the LOS velocity decreases more rapidly away
from the center of the CME, changing the spectral morphology from the
semicircular pattern of Case 1 to a more pointed structure.

The fourth configuration combines the GCS-constrained CME orientation and tilt
with the asymmetric lateral expansion of Case 3. It therefore retains
the Doppler asymmetry and displacement produced by the CME  axis tilt while producing the pointed spectral morphology associated with
asymmetric lateral expansion. Among the four configurations, Case 4
shows the closest qualitative correspondence with the VELC
observations, both in the spectral morphology and its temporal
evolution.

These configurations are intended to isolate the geometric effects that
influence the synthetic spectra. They are not a parameter search or a
quantitative inversion of the observed CME geometry or velocity field.
A quantitative inversion framework will be developed in future work to
constrain the three-dimensional CME geometry and velocity field from
spectroscopic observations.

\section{GCS Fitting}
\label{app:gcs}

We used the Graduated Cylindrical Shell (GCS) model
\citep{2006ApJ...652..763T,2009SoPh..256..111T,2011ApJS..194...33T}
to estimate the propagation direction and axis orientation of the CME.
The reconstruction was performed using multi-viewpoint coronagraph
observations from SOHO/LASCO and STEREO/COR, together with the
available EUV observations (Fig.~\ref{fig:fig_a2}). The GCS wireframe was adjusted to the observed CME morphology in the different viewpoints.

The best-fit configuration gives a CME longitude of $97^\circ$,
latitude of $-8^\circ$, and an axis tilt of $54^\circ$. These parameters
are used in our three-dimensional CME model to set the CME
orientation with respect to the observer. They are not adjusted during
the spectral modeling and are not used to derive the CME velocity
field. The GCS reconstruction therefore, provides the orientation
constraint for the model, while the lateral expansion
parameters are varied independently.

\bibliographystyle{aasjournalv7}

\end{document}